\newif\ifaastex\aastexfalse

\ifaastex
\documentclass[manuscript]{aastex}
\else
\documentclass{emulateapj}
\fi

\usepackage{natbib}
\ifaastex
\else
\let\jjdagger=\dagger
\renewcommand{\dagger}{\ensuremath{\jjdagger}}
\fi

\newcommand{\halpha}{H$\alpha$}
\newcommand{\lyb}{Ly$\beta$}
\newcommand{\kms}{\ensuremath{\mathrm{km\,s}^{-1}}}
\newcommand{\cmthree}{\ensuremath{\mathrm{cm}^{-3}}}

\newcommand{\hms}[3]{#1$^{\mathrm h}$ #2$^{\mathrm m}$ #3$^{\mathrm s}$}
\newcommand{\dms}[3]{#1$\arcdeg$ #2$\arcmin$ #3$\arcsec$}

\shorttitle{HST Observation of Tycho}
\shortauthors{Lee et al.}

\begin{document}

\title{Resolved Shock Structure of the Balmer-Dominated
  Filaments in Tycho's Supernova Remnant : Cosmic Ray Precursor?
}

\author{Jae-Joon Lee,\altaffilmark{1,2}, 
John C. Raymond\altaffilmark{3}, 
Sangwook Park\altaffilmark{1},
William P. Blair\altaffilmark{4},
Parviz Ghavamian\altaffilmark{5},
P. F. Winkler\altaffilmark{6},
Kelly Korreck\altaffilmark{3},
}

\altaffiltext{1}{Astronomy and Astrophysics Department, Pennsylvania
  State University, University Park, PA 16802, USA}
\altaffiltext{2}{lee@astro.psu.edu}
\altaffiltext{3}{Harvard-Smithsonian Center for Astrophysics, 60
  Garden Street, Cambridge, MA 02138, USA}
\altaffiltext{4}{Department of Physics and Astronomy, Johns Hopkins
  University, 3400 N. Charles St., Baltimore, MD 21218, USA}
\altaffiltext{5}{Space Telescope Science Institute, 3700 San Martin
  Drive, Baltimore, MD, 21218, USA}
\altaffiltext{6}{Department of Physics, Middlebury College,
  Middlebury, VT 05753, USA}

\begin{abstract}
  We report on the results from \halpha\ imaging observations of the
  eastern limb of Tycho's supernova remnant (SN1572) using the
  Wide Field Planetary Camera-2 on the Hubble Space Telescope. We
  resolve the detailed structure of the fast, collisionless shock wave
  into a delicate structure of nearly edge-on filaments.
  %We derive preshock
  %conditions for these shocks using crosscuts perpendicular to the
  %local shock fronts and an Ha emission model. 
  We find a gradual increase of \halpha\ intensity just ahead of the
  shock front, which we interpret as emission from the thin
  ($\sim1\arcsec$) shock precursor. We find that a significant amount
  of the \halpha\ emission comes from the precursor and that this
  could affect the amount of temperature equilibration derived from
  the observed flux ratio of the broad and narrow \halpha\ components.
  The observed \halpha\ emission profiles are fit using simple
  precursor models, and we discuss the relevant parameters. We suggest
  that the precursor is likely due to cosmic rays and discuss the
  efficiency of cosmic ray acceleration at this position.
\end{abstract}

%\keywords{ISM:supernova remnants -- ISM: individual (Tycho,
%  G120.1+1.4) -- Shock Waves -- line: profiles}


\section{INTRODUCTION}
\label{sec:intro}

% [Particle acceleration in shocks]

The shock transition in fast astrophysical shocks is intrinsically a
``collisionless'' process, and energy is dissipated via plasma
turbulence and/or electromagnetic fields.  An important consequence of
the collisionless nature of the shocks 
is cosmic ray acceleration \citep[e.g.,][]{1987PhR...154....1B}.
%is that
% As a result, these shocks do not partition
%energy equally among different particle species, and 
%they accelerate cosmic rays .
While there is increasing evidence of cosmic ray acceleration in
supernova remnants, the details of the process are still not well
understood, and the question of whether supernova remnants (SNRs) are
the primary acceleration sites of Galactic cosmic rays is still
open \citep{2009Natur.460..701B}.

% and the electrons that produce radio and X-ray synchrotron
%emission. 

% However, the efficiency of cosmic ray acceleration is poorly
% understood. Moreover, most observations of shocked gas are sensitive
% only to the electron temperature, so that the electron-proton
% temperature ratio, Te=Tp, must be understood to interpret observations
% correctly. 

Cosmic ray acceleration models require a precursor in which
accelerated particles can scatter back to the postshock region for
further acceleration. 
% and forth between the shock jump and the
%upstream turbulence. 
Observations of the cosmic ray precursor can constrain the two
key parameters of acceleration models; the diffusion coefficient and
the injection efficiency
\citep{1987PhR...154....1B,1988ApJ...333..198B}.
% , thus provides
% opportunity to address the details of acceleration process in the
% shock.
The Balmer-dominated filaments that are produced when fast SNR shocks
propagate into partially neutral gas are potential sites where such a
cosmic ray precursors can be observed. Most of the Balmer emission
comes from a very narrow zone behind the shock, where the hydrogen
atoms swept up by the shock are excited before they are ionized
\citep{1978ApJ...225L..27C,1980ApJ...235..186C}.  However, in the
presence of the precursor, additional Balmer emission is expected from
the precursor region where the preshock gas is compressed and heated.
%The excitation of the neutral hydrogens in the CR precursor,
%where the preshock gas is heated, 
% is another source of Balmer
% emission. 
% Therefore, the observation of \halpha\ emission from the CR
% precursor can provide an unique opportunity study the structure of the
% CR precursor.
Using long-slit \halpha\ spectroscopy along the shock normal of a
Balmer filament in Tycho's SNR, \citet[][Lee07
hereafter]{2007ApJ...659L.133L} found that there is an increase of the
\halpha\ narrow component intensity in a small region ($\sim
0.4\arcsec$) ahead of the shock front, which they proposed as
potential emission from the precursor. However, the angular resolution
of the observation by Lee07 is $\sim 0.5\arcsec$ and their results
needed to be verified with high resolution observations.

In this \emph{Letter}, we report \halpha\ imaging observations of
Balmer-dominated filaments in Tycho's SNR using the Wide Field and Planetary
Camera 2 (WFPC2) on the Hubble Space Telescope (HST), which resolves
the detailed structure of the shock. \S~\ref{sec:observations}
presents the observations and reports the detection of the
precursor. The precursor is modeled in \S~\ref{sec:model} and its
characteristics are discussed in \S~\ref{sec:analysis}. Finally in
\S~\ref{sec:summary}, we discuss the efficiency of cosmic ray
acceleration in this region.

\begin{figure}[b]
  \plotone{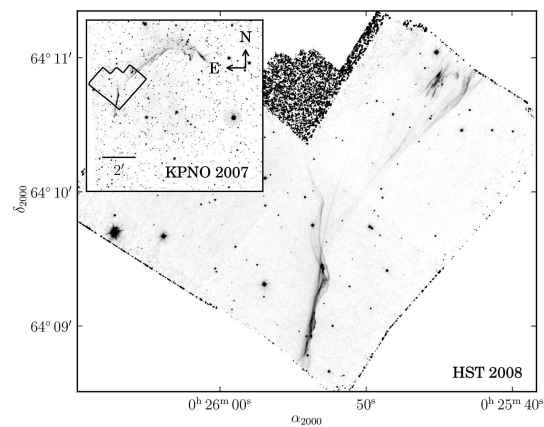}
  \label{fig:kpno-hst}
  \caption{ Hubble Space Telescope image of
    the Balmer-dominated filaments in the northeastern part of
    Tycho's SNR.  The WFPC2 detector with F656N filter was used. The inset
    shows an \halpha\ image toward Tycho's SNR from
    a ground-based telescope (KPNO). The area observed by HST is marked.
}
\end{figure}

\section{Observations and Results}
\label{sec:observations}

We observed the eastern limb of Tycho's SNR using WFPC2 on HST.
%on board the
%Hubble Space Telescope (HST).
%\emph{HST}'s Wide Field and Planetary Camera 2 (WFPC2) imaged
%northeastern part of Tycho. WF3 camera is centered at brightest
The observations were conducted on Mar 23, 2008, with the camera
arranged to cover knot g, one of the brightest
\halpha\ filaments in the remnant ($\alpha_{2000}$,
$\delta_{2000}$=\hms{00}{25}{52}, \dms{+64}{09}{21}).
% \footnote{We note
%   that the \halpha\ emission in this region changed since the original
%   observation of REFERENCE and the knot g actually has faded away.}.
% at coordinates of (00 25 52, +64 09 21).  
A total of 10 exposures, each with $2600\sim2700$ sec, were
obtained % at 2008-03-23,
%resulting total of  15800 (?) sec exposure. 
using the F656N \halpha\ filter, which has a central wavelength
of %is centered
6564\AA. 
%\ and a bandwidth of 54\ \AA\ ($\sim 2500\,\kms$).  
The filter has a bandwidth of 54~\AA\ ($\sim 2500\,\kms$),
which is comparable to the velocity width of the
\halpha\ broad component in this region \citep{2001ApJ...547..995G},
and transmits 
%has a velocity width comparable to the filter bandwidth, and 
about  half of the broad component flux.
%this filter passes all of the narrow
%component and, based on the spectrum of Ghavamian et al., about 1/2
%(?) of the broad component.
%F656N centered on 6563.758, BW=53.768
%15800s (~2700*6???)
Images are combined and ``drizzled'' onto a $0.06\arcsec$ pixel$^{-1}$
scale using the IRAF task \emph{multidrizzle}
\citep{2002PASP..114..144F}, which also detects and removes
cosmic-rays.
%, improving the sampling of the
%WF. 
The PSF of the final drizzled image has a FWHM of $\sim 0.18\arcsec$.

\begin{figure}
  \plotone{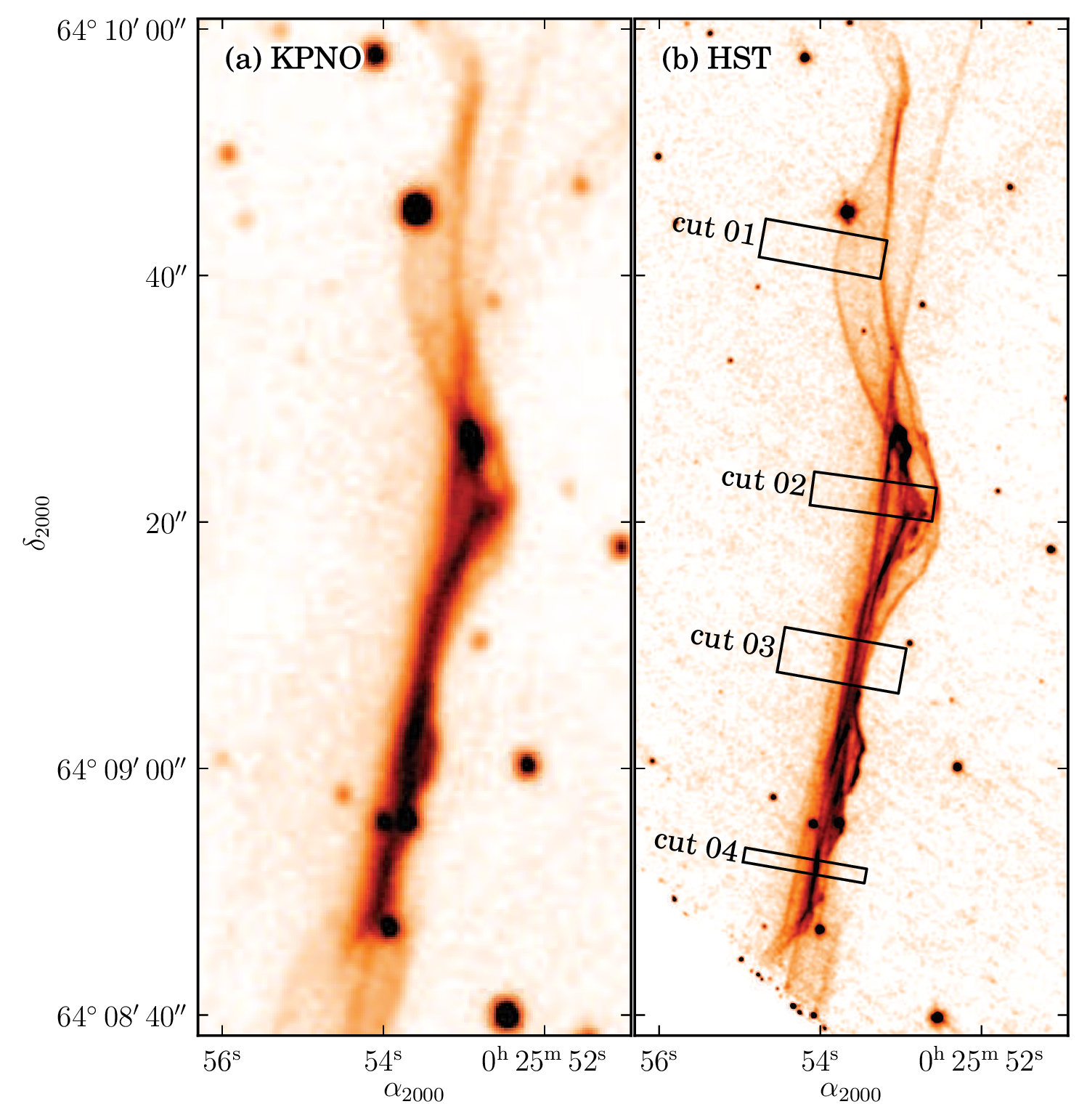}
  \caption{ (a) The \halpha\ image of knot g, 
    one of the brightest  Balmer-dominated filaments in Tycho, taken
    by KPNO 2.1m telescope. (b) The HST image of the same region. The black
    rectangles are 
   regions from which the
    brightness profiles are extracted. 
    \label{fig:hst-kotg}}
\end{figure}

\begin{figure}
  \plotone{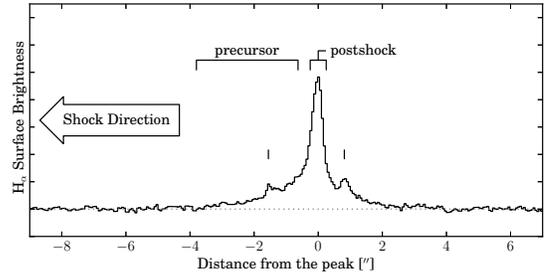}
  \caption{\halpha\ brightness profile across the shock normal
    extracted from cut~04. The bright narrow peak is believed to
    correspond to the emission from the narrow region of the immediate
    postshock area. 
    The emission extends up to 4\arcsec\ toward upstream, which is
    interpreted as emission from the shock precursor. The small bumps
    in the profile (marked with small vertical bars) are likely the
    projection of other shock fronts.
    \label{fig:profile-southe}}
\end{figure}

% In Fig.~\ref{fig:kpno-hst}(a), we show the \halpha\ image of Tycho
% observed with a ground-based telescope (CCD image obtained at the KPNO
% 2.1m telescope on 2007 October 4), and overlay the field-of-view of
% the HST observations. 
The full field observed with HST is shown in Fig.~\ref{fig:kpno-hst}.
The brightest filaments, comprising the knot g region and associated
filaments, are seen on the WF3 chip, while fainter filaments
belonging to the northeastern limb of the SNR are visible on the WF4
chip.
% overlaid on
%the ground-based \halpha\ image of Tycho, 
%together with the observed
%HST image. 
The superb angular resolution of HST reveals details of the filaments
not available from ground-based telescopes. This is more clearly seen
in Fig.~\ref{fig:hst-kotg}, where we compare the close-up view of the
knot g region to the image observed with a ground-based
telescope (CCD image obtained at the KPNO 2.1m telescope on 2007
October 4) at the same scale.
% shows zoomed-up view of the region
%around the knot g. 
%, much
%complicated than what HST has revealed for the Balamer-dominated
%shocks in Cygnus loop and SN1006
%\citep{1999AJ....118..942B,2007ApJ...659.1257R}.  Despite the
%complicated shock structures in the region, % region is complicated,
% with projection of multiple shocks,
%with also complicated structure for each one of the shocks, 
The HST image reveals a faint extension of the emission toward
upstream (to the east) along most of the bright part of the filaments
(between $\delta$=\dms{64}{08}{40} and $\delta$=\dms{64}{09}{20}).
%down from $\delta$=\dms{64}{08}{40} up to $\delta$=\dms{64}{09}{20}.
% The bright think filaments are believed to represent the emission
% from the post shock area (REFERENCE).
This extension is more clearly demonstrated by the \halpha\ brightness
profiles from the cuts along the shock normals, as seen in
Fig.~\ref{fig:profile-southe}. The profiles show a bright emission
peak of thickness $\lesssim 0.5\arcsec$,
%thin filament of
%FWHM$\lesssim 0.5\arcsec$, 
which is the emission from neutral hydrogen excited in the postshock
area. The small bumps around offsets $-1.8\arcsec$ and $0.7\arcsec$
are likely due to the projection of fainter tangencies of the rippled
shock front to the line of sight.
%overlapping shocks.  
The \halpha\ emission slowly falls off not only toward the downstream
but also toward the upstream direction.  In the downstream, the
\halpha\ emission is emitted within a very narrow region behind the
shock as neutral hydrogen is rapidly ionized.  Thus, we consider that
the downstream emission is likely a projection of the curved shock
fronts.  Some of the upstream emission could be in principle
attributed to a similar projection effect. However,
Fig.~\ref{fig:hst-kotg} shows that the shock front in this region does
not show any significant curvature, and it is difficult to devise a
shock geometry that explains both upstream and downstream emission.
Fig.~\ref{fig:simple-fit} shows emission profiles along different
cuts, and the upstream emission components are similarly seen.  The
upstream emission is also seen faintly in the profile of cut~01 where
the shock surface is apparently convex (see Fig.~\ref{fig:hst-kotg}),
and a projection effect is not likely to explain the upstream
emission.
% shows similar increase of the
% intensity as observed in other shocks. However, the increase of the
% emission seems to occur over a region relatively larger than other
% shocks, and the nature of the upstream emission in cut~01 is not
% conclusive.
%On the other hand, 
Therefore, we propose that the faint emission from the upstream region
represents the emission mostly from the neutral hydrogen atoms excited
in a shock precursor, and the effect of shock geometry makes only a
minor contribution.

\section{Modeling the Precursor Emission}
\label{sec:model}

\begin{figure}
  \plotone{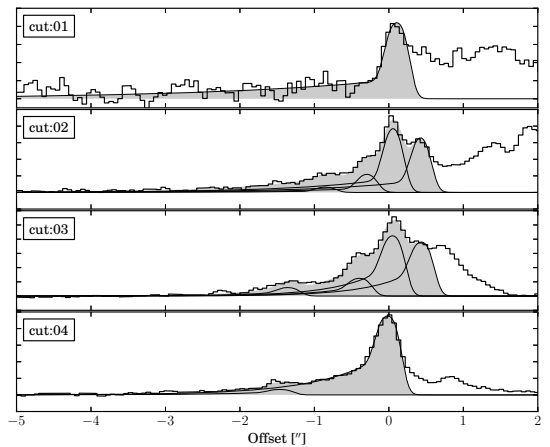}
  \caption{\halpha\ brightness profiles across the shock normals in
    different regions marked in Fig.~\ref{fig:hst-kotg}. The profiles
    are fitted with multiple components of toy precursor models. The
    gray area represents the accumulated emission from all models, and
    the solid lines represent emission from projected individual shocks.
    \label{fig:simple-fit}}
\end{figure}

To estimate the physical properties of the precursor, we model the
observed spatial \halpha\ emission profiles
(Fig.~\ref{fig:simple-fit}).  We first try a simple toy precursor
model.  We assume a shock front located at $x=x_0$ with increasing $x$
toward downstream. The \halpha\ emission from the postshock region is
assumed to be confined to a region of thickness of $w$ with a uniform
emissivity.  In the precursor, the \halpha\ emissivity peaks at the
shock front and decreases exponentially away from the shock front.
%in the precursor increase exponentially, peaking just
%before the shock front. 
The profile is written as
\[y = \left\{ 
\begin{array}{l l}
  0 & \quad \mbox{if $x - x_0 > w$}\\ 
  Fw^{-1} & \quad \mbox{if $0 \le x-x_0 \le w$}\\ 
  FfL^{-1} e^{(x-x_0)/L} & \quad \mbox{if $x-x_0 < 0$}\\
\end{array} \right. \]
where $F$ is the total \halpha\ flux from the postshock region, $L$ is
the precursor length scale, and $f$ is a flux ratio of the precursor
to the postshock region.
%the fraction of the postshock contribution.
The variation in the downstream region could be modeled with a curved
geometry of the shock.  However, we find that the observed variation
cannot be adequately fit by shocks of a simple geometry.
%uniform surface brightness. 
Instead of introducing arbitrary fit parameters to describe the
structure along the line of sight, we simply assume plane-parallel
shocks and ignore the data where the model deviates from the
observation.
%in the slow fall-off in the postshock. % area are ignored in the fit.
% (e.g., data included in the fit is indicated in
%Fig.~XXX). 
As long as $L$ is sufficiently smaller than the local curvature radii
of the shocks, the plane-parallel assumption will not significantly
affect estimated precursor parameters.
%This may overestimate $L$ and also affect other parameters,
%but given that $L$ is significantly smaller than the local shock
%curvature, the effect would not be significant.
%The possible consequences of this assumption are discussed later.  
As evident from Fig.~\ref{fig:hst-kotg}, the profiles require multiple
shock components projected along the line of sight.
%is adequately fit by a
%single shock and multiple components are required. 
To minimize the number of free parameters, we assume that shocks have
the same profile shape, i.e., parameters $f$, $L$ and $w$ are tied
among multiple shocks for a given profile cut from
Fig.~\ref{fig:hst-kotg}.  The model profiles are Gaussian smoothed to
account for the instrumental profile.

\newcommand{\TableOneCaption}{Fit parameters from the toy precursor model}
\newcommand{\fitresult}[3]{{ #1$_{#2}^{#3}$}}

%\ifaastex
\begin{deluxetable}{ccccc}
%\else
%\begin{deluxetable}{ccccc}
%\fi
\tablecolumns{4}
\tablewidth{0pc}
\tablecaption{\TableOneCaption
}
\tablehead{
\colhead{} & \colhead{$f$} & \colhead{$L$ [\arcsec]} & \colhead{$w$ [\arcsec]} & \colhead{$\chi^2$/d.o.f.}
}
\startdata
{cut 01} 
  & \fitresult{1.4}{-0.2}{+0.3}
  & \fitresult{2.5}{-0.5}{+0.7}
  & \fitresult{0.42}{-0.05}{+0.06}
  & 206/170\\

{cut 02} 
  & \fitresult{0.72}{-0.06}{+0.06}
  & \fitresult{1.2}{-0.1}{+0.1}
  & \fitresult{0.31}{-0.02}{+0.02}
  & 170/182\\

{cut 03} 
  & \fitresult{0.57}{-0.03}{+0.04}
  & \fitresult{0.70}{-0.05}{+0.06}
  & \fitresult{0.35}{-0.01}{+0.01}
  & 258/182\\

{cut 04} 
  & \fitresult{0.93}{-0.03}{+0.04}
  & \fitresult{0.87}{-0.05}{+0.05}
  & \fitresult{0.39}{-0.02}{+0.02}
  & 220/182
\enddata
%\tablecomments{Table comments.} 
%\tablenotetext{\dagger}{The temperature and the
%  ionization time scale is fixed at the best fit values of XN}
%\ifaastex
\end{deluxetable}
%\else
%\end{deluxetable}
%\fi

%\input{tab1}

The fits are shown in Fig.~\ref{fig:simple-fit}, and the results are
summarized in Table~1.  We find that the \halpha\ flux from the
precursor region is comparable to that of the postshock area (f $\sim
1$), with $L$ around 1\arcsec, corresponding to $3\times10^{16}$~cm at
the assumed distance of 2.1 kpc to Tycho
\citep{1978ApJ...224..851K,2001ApJ...547..995G}.
%The
%fraction of \halpha\ emission from the precursor ($f$) is indeed
%significant, $0.5 \sim 1.$. 
The thicknesses of the postshock emitting area are around $\sim 0.3
\arcsec$. A relatively large $L$ of $\sim 2\arcsec$ is found in
cut~01, but this is likely to be overestimated due to the much
stronger local curvature of the shock in this region.  We also note
that, for profiles in cuts~02 and 03, where the structure of the
overlapping shocks is quite complex, the fitted precursor parameters
are somewhat sensitive to the assumed baseline and also the number of
shock components.  These can cause $\sim20-30 \%$ systematic
uncertainties on the fitted parameters.  % We find that this can lead to
% uncertainties of a few tens of percent of the fitted parameters.

We now consider a more realistic precursor model that assumes an
exponential temperature profile (similar to the above toy model) in
the precursor.  The model calculates the ionization of hydrogen atoms
throughout the shock and the emissivity of \halpha\ and \lyb\ lines.
The radiative transfer of the \lyb\ line is computed using the Monte
Carlo technique to account for the \halpha\ enhancement by
Ly$\beta$-trapping.  The model has been utilized by
\citet{2009ApJ...690.1412W} to interpret the result of Lee07. We adopt
the parameters used in \citet{2009ApJ...690.1412W}; shock velocity of
2000 \kms, preshock density of 1 \cmthree, and the preshock neutral
fraction of 0.85.  More details of the model and the input parameters
can be found in \citet{2009ApJ...690.1412W}. The extensive discussion
of the detailed modeling and associated uncertainties is beyond the
scope of this \emph{Letter}, and here we simply present a brief summary of
the results.
% Given the uncertainties
% of the input parameters (e.g., shock velocity, preshock density,
% preshock neutral fraction, etc.) and their possible variation along
% the filaments, we only provide a brief summary of the results instead
% of discussing the details.  
For cuts 02, 03, and 04, we estimate peak
temperatures in the precursor T$_{\mathrm{peak}} = 80,000 \sim
100,000$ K, and the precursor length scale $L= 5 \sim 7\times10^{16}$
cm.
%, and $w=0.8 \sim 1.2 \times10^{6}$
%cm. 
As in the toy model, a larger length scale is required for cut 01.
% , but the result would be
% vulnerable to the assumed geometry.  
The estimated length scales of the precursor are generally larger than
those estimated from the simple toy model. This is because the
\halpha\ emissivity is sensitive to the temperature, i.e., the
emissivity profile increases more rapidly than the temperature
profile, thus effectively reducing the length scale. We note that the
precursor parameters are in agreement with the results of
\citet{2009ApJ...690.1412W}, while the length scale is slightly
larger.
%by a few tens of percent.

% could be partially responsible for the observed width, it may not be
% enough the dominant contributor.

% The curvature 
% of the shock could be responsible.
% We attribute
% the observed broadness of the filaments to the curvature of the shock
% and/or the small scale density fluctuation of the medium. For the
% profile fitting.

\section{Origin of the Precursor}
\label{sec:analysis}

The existence of a thin precursor has been suggested from previous
observations.
%In most of the Balmer filaments studied at high spectral resolution,
%
The spectral profiles of the \halpha\ narrow component traces the
velocity distribution of the gas entering the shocks, and those
observed in SNRs have widths of 40 -- 60 \kms\ \citep[e.g.,
][]{1994ApJ...420..286S,2003A&A...407..249S}. This is too large for
the temperature of the ambient gas, as all the hydrogen would have
been ionized at the implied temperature and no Balmer filament should
exist. Instead, the observed line width is suggested to represent the
gas heated in the precursor, which is thin enough for the preshock
neutrals not to be completely ionized.
%, but thick enough
%to be heated and accelerated.  
Also, the observed flux ratio of the \halpha\ broad component and the
narrow component was sometimes found to be smaller than what models predict
\citep[e.g,][]{2003ApJ...590..833G,2009ApJ...696.2195R}, and
the excessive narrow component emission was attributed to the
contribution of emission from the precursor
\citep{2001ApJ...547..995G,2009ApJ...696.2195R}.

% Previous spectroscopic observations of various Balmer-dominated
% filaments indeed found that the observed ratio of the narrow component
% flux to the broad component one is noticeably higher than 

The characteristics of the precursor revealed by our HST observations
are consistent with results from previous observations. The peak
temperature in the precursor may be relatively higher than the
temperature implied by the line width of the \halpha\ narrow
component. However, the temperature we modeled is the electron
temperature, which might not be in equilibrium with the neutral
hydrogen atoms. Also, the velocity profiles of the \halpha\ emission
could deviate from a Gaussian profile
\citep{2001ApJ...547..995G,2007ApJ...659L.133L,raymond2010}, and a
simple line width may not be an adequate temperature indicator.  Our
observations show that the emission from the precursor is a
significant contributor to the \halpha\ narrow component.  We estimate
that the precursor emission may contribute up to 30-40\% of the narrow
component for a slit width of $1\arcsec$. The narrow component
emission from the precursor will affect estimates of electron-ion
equilibration based on the observed broad-to-narrow intensity ratio
\citep{2001ApJ...547..995G,2007ApJ...654..923H,2009ApJ...696.2195R}.
For those SNRs where the observed \halpha\ broad-to-narrow flux ratio
was smaller than the model predictions, accounting for the precursor
could bring them into agreement.

The likely candidates for producing this precursor are a cosmic ray
precursor or a fast neutral precursor
\citep{1994ApJ...420..286S,1994ApJ...420..721H}.  While our HST
results provide detailed structure of the precursor, models of these
precursors are not available for a quantitative comparison.
%Unfortunately, however, our HST results are not very useful to
%distinguish between two models as detailed quantitative models of the
%precursor structure are generally not available for both scenarios.
However, the cosmic ray precursor scenario
has been preferred over the fast neutral precursor for various reasons
(see Lee07 and references therein).
%While the discussion of the nature of the precursor have been limited
%for qualitative ones (e.g., Lee07), .  
Also, the growing evidence of cosmic ray
acceleration in Tycho's SNR supports the existence of cosmic ray
precursor \citep[e.g.,][]{2005ApJ...634..376W}.
% , makes the cosmic ray precursor as a more natural scenario.
% Nevertheless. our inferred characteristics of the precursor could be
% used to constrain the cosmic ray acceleration models.

\citet{2009ApJ...690.1412W} computed a series of time dependent
numerical simulations of cosmic ray modified shocks, tuning the model
parameters to reproduce the precursor characteristics of Lee07.  
%, by and large, still applicable to our new results 
The results of \citet{2009ApJ...690.1412W} can be a plausible
approximation as the estimated precursor properties are not much
different.  They found that, assuming a distance of 2.1 kpc to Tycho's
SNR, the CR diffusion coefficient, $\kappa$, the injection parameter,
$\epsilon$, and the timescale for the energy transfer, $\tau$, of
$\kappa = 2 \times 10^{24}$ cm$^2$ s$^{-1}$, $\epsilon = 4.2 \times
10^{-3}$, and $\tau = 426$ yr is required to describe the
observations.  The length scales of the precursor estimated from our
new observation are slightly larger than that of Lee07. The larger
precursor length scale requires a larger cosmic ray diffusion
coefficient. This increases the acceleration time scale, so the cosmic
ray injection parameter may need to be increased to compensate for the
slower acceleration.
% time scale for energy transfer from the cosmic ray to precursor
% should increase.
While some fine tuning of the parameters may be required, we believe
that \citeauthor{2009ApJ...690.1412W}'s findings, e.g., that the
cosmic ray acceleration is not very efficient in this shock and about
10\% of the shock energy has converted to cosmic rays, are still
valid.

\section{Cosmic ray Ion Acceleration in Tycho}
\label{sec:summary}

The existence of the cosmic ray precursor 
%imply the existence of
%cosmic ray acceleration in the shock, but it 
does not necessarily imply efficient cosmic ray acceleration. For the
Balmer-dominated filaments to be observable, some neutral hydrogen
needs to survive ionization in the precursor. As efficient cosmic ray
acceleration tends to make the precursor wider and hotter, the
Balmer-dominated filaments may not trace the shocks having efficient
cosmic ray acceleration \citep[cf.][]{2009Sci...325..719H}.
% so that the preshock neutrals
% hydrogens will be completely ionized in the precursor.  On the other
% hand, the existence of the \halpha\ broad component imply that the
% preshock neutral hydrogens indeed reaches the shock.
For the shock studied in this paper, a more direct suggestion of
inefficient acceleration can be inferred from the results of Lee07.
%On the other hand, Lee07 estimated
%that 
From the difference in radial velocity of the preshock gas and the
\halpha\ narrow component emitted in the postshock region, they
estimated the amount of gas deceleration in the precursor. The preshock
gas is decelerated in the precursor due to the gradient of the cosmic
ray pressure by about a few hundred \kms\ based on the radial
velocity measurements. The value is not sensitive to the assumed
distance, and is significantly smaller than the measured line width
of the broad component.  %The result is interpreted that, 
In this
shock, the thermal pressure of the ordinary gas still dominates over
the cosmic ray pressure and the acceleration is not likely to be
very efficient.

Throughout the discussion, we have assumed that the distance to
Tycho's SNR is 2.1 kpc. The measurement is based on the estimated
proton temperature (from the observed line width of the broad
component) and the optical proper motion, assuming no cosmic ray
acceleration at the shock. The cosmic ray acceleration, if
significant, can effectively reduce the postshock temperature
\citep[see][and references therein]{2009Sci...325..719H}; thus the
distance may have been underestimated. However, as has been discussed
above, the optical observations are consistent with cosmic ray
acceleration not being very efficient in this region, so the distance
of 2.1 kpc, derived assuming no cosmic ray acceleration, remains
a reasonable value.

Using Chandra observations, \citet{2005ApJ...634..376W} found the
locations of the shock front (SF) and the contact discontinuity (CD)
along the boundary of the remnant. They interpreted the small
separation between the two as an indication of efficient cosmic
ray acceleration.  The region of knot g is one of the regions
where the SF-CD separation is smallest (except those regions of ejecta
protrusion). A simple extrapolation of their argument will lead to the
most efficient cosmic ray acceleration in this region, being
inconsistent with our results.  However, the region around knot g is
where the remnant could be interacting with dense ambient clouds
\citep{1997ApJ...491..816R,2004ApJ...605L.113L}.  Thus, the small SF-CD
separation in this region could be due to a recent encounter of the
shock with the
dense ambient gas, instead of efficient cosmic ray acceleration.

In conclusion, we have presented high resolution \halpha\ imaging
observations of Tycho, revealing the existence of a thin precursor
which we interpret as a cosmic ray precursor.  While the current
observation is consistent with
%suggest that 
inefficient acceleration,
%the acceleration being not very efficient, 
the observation of the precursor itself provides an important
opportunity to constrain the key parameters of the acceleration, such
as the diffusion coefficient and the injection parameters. A
comparison with detailed numerical simulations will be critical to
study the detailed physics of cosmic ray acceleration.

\acknowledgements
This research was supported by STScI grant GO-11184.01-A-R to the Smithsonian
Astrophysical Observatory.

% Therefore, the current observations seems consistent with
% the distance of 2.3 kpc.

{\it Facilities:}  \facility{HST (WFPC2)},  \facility{KPNO2m}

%\bibliography{ms}

\end{document}